\documentclass{article}

     \PassOptionsToPackage{numbers, compress}{natbib}

\usepackage[preprint]{neurips_2026}

\usepackage[utf8]{inputenc} %
\usepackage[T1]{fontenc}    %
\usepackage{hyperref}       %
\usepackage{url}            %
\usepackage{booktabs}       %
\usepackage{amsfonts}       %
\usepackage{nicefrac}       %
\usepackage{microtype}      %
\usepackage{xcolor}         %
\usepackage{graphicx}
\graphicspath{{figures/}}

\usepackage{amsmath}
\usepackage{amssymb}
\usepackage{algorithm}
\usepackage{algpseudocode}
\usepackage{subcaption}

\title{Energy-based models for diagnostic reconstruction and analysis in a laboratory plasma device}

\author{%
  Phil Travis\thanks{Work performed as part of his PhD thesis at the UCLA Department of Physics and Astronomy} \\
  Ergodic LLC, USA \\ 
  \texttt{phil@physicistphil.com} \\
  \And
  Troy Carter \\
  Oak Ridge National Laboratory, Oak Ridge, Tennessee 37830, USA \\
  \texttt{carterta@ornl.gov} \\
}

\begin{document}

\maketitle

\begin{abstract}
  Energy-based models (EBMs) provide a powerful and flexible way of learning a joint probability distribution over data by constructing an energy surface. This energy surface enables insight extraction and conditional sampling. We apply EBMs to laboratory plasma physics, a domain characterized by highly nonlinear phenomena. These phenomena are studied using plasma diagnostics, which are often difficult to analyze and subject to hardware degradation. In addition, the possible configuration space of a plasma device is sufficiently large that it cannot be efficiently searched using conventional analysis techniques. EBMs address these issues. At the Large Plasma Device (LAPD), a CNN- and attention-based EBM is trained on a set of randomly generated machine conditions and their corresponding diagnostic time series. We demonstrate diagnostic reconstruction using this EBM on real data and show that additional diagnostics improves reconstruction error and generation quality. The energy surface is directly evaluated for an ill-posed inverse problem: inferring probe position from a time-series measurement. This inference illuminates symmetries in the data, potentially leading to a method of inquiry to supplement conventional data analysis. Trends in diagnostic signals are inferred via conditional sampling over machine inputs. In addition, this multimodal EBM is able to unconditionally reproduce all distributional modes, suggesting future potential in anomaly detection on the LAPD. Fundamentally, this work demonstrates the flexibility and efficacy of EBM-based generative modeling of laboratory plasma data, and showcases multiple practical uses of just a single trained EBM in the physical sciences. 
\end{abstract}

\section{Introduction: goals and prior work}

\subsection{Goals}
We seek to use machine learning -- particularly generative models -- to alleviate some of the challenges facing fusion-related plasma science and to accelerate the advent of fusion power. In particular, we wish to reconstruct missing diagnostic signals or signals elsewhere in the plasma from a smaller set of data to improve experimental efficiency. In addition, we seek combine our data with other sources of information such as simulations or constraints placed on it via prior knowledge. We also want to extract maximal insight from data collected. These tasks are a natural fit for generative modeling.

\subsection{Generative modeling of plasmas}

The use of generative models in plasma physics is not without precedent, but remains relatively uncommon. Variational autoencoders (VAEs) \cite{kingma_auto-encoding_2022} have been used for relationship discovery \cite{lennart_van_rijn_minimizing_2022, vos_discovery_2024} and mode or state identification \cite{skvara_detection_2020, wei_dimensionality_2021} in a various fusion devices. Generative-adversarial networks (GANs) \cite{goodfellow_generative_2014} have also been used to generate synthetic data for training classifiers \cite{dave_synthetic_2023} and inverse problems \cite{juven_generative_2024}. Machine learning applications to plasma physics has been recently reviewed \cite{pavone_machine_2023}.

Directly relevant to this work, diagnostic reconstruction has been used to upsample Thompson scattering signals \cite{jalalvand_multimodal_2024}, and randomized experiments were used to train a VAE as a surrogate for a collisional radiative model \cite{daly_data-driven_2023}. Diversified experiments have also been used for profile prediction and optimization in the Large Plasma Device \cite{travis_machine-learned_2025}. Concerning EBMs, they have yet to be applied to plasma physics problems. One notable application in the physical sciences has been in the high-energy physics community: EBMs were used for modeling event patterns in the Large Hadron Collider (LHC) for anomaly detection and to augment a classifier \cite{cheng_versatile_2024} with success.

\subsection{Introduction to energy-based models (EBMs)}

Energy-based models interpret a probability distribution through the lens of the Boltzmann distribution \cite{hopfield_neural_1982, ackley_learning_1985, lecun_tutorial_2006}. In the EBM formulation, the unnormalized probability density is parameterized by an energy function, that is $\tilde{p}(\mathbf{x}) = \exp(-E_\theta(\mathbf{x}))$, where $\theta$ are the parameters of this energy function. In this work and the works cited, this energy function is parameterized by a neural network. EBMs have been historically difficult to train, but recent work has demonstrated high-quality sampling using MCMC techniques in high-dimensional spaces \cite{du_model_2019, du_implicit_2020, du_improved_2021, du_compositional_2020, du_unsupervised_2021, nijkamp_anatomy_2020, nijkamp_learning_2019, deng_residual_2020, gao_learning_2018}. These MCMC techniques are fundamentally based in contrastive divergence \cite{hinton_training_2002, ruslan_deep_2009, tieleman_training_2008}. The nature of MCMC-based sampling of EBMs, detailing the convergence of and expansion and contraction of learned models (a paradigm particularly helpful for training EBMs in this work) has been studied \cite{nijkamp_anatomy_2020, nijkamp_learning_2019}. 

EBMs can attain GAN-like performance while generating all modes of a distribution \cite{du_implicit_2020}. These models have also been used for text generation \cite{deng_residual_2020} and model-based planning \cite{du_model_2019}. In addition, EBMs can be composed by combining the energy functions in various ways which has been demonstrated in image generation tasks \cite{du_compositional_2020, du_unsupervised_2021}. EBMs can be competitive with more commonly-used generative models \cite{carbone_hitchhikers_2024, bond-taylor_deep_2021}.

EBMs in particular have some desirable properties. The implicit generation scheme does not require balancing an explicit generator with a discriminator or encoder. In addition, the energy-based representation is amenable to modification of an auxiliary function, allowing the generated samples to be easily steered, enabling the composition of multiple energy functions. Lastly, EBMs do not typically suffer from mode collapse. For these reasons we use EBMs for building a distribution of data from the LAPD.

\section{Data source and preparation}
The Large Plasma Device \cite{LAPD_LaB6} is a basic plasma science device capable of generating magnetized plasmas up to 18m long and 1m in diameter. The plasma is formed by a thermionic cathode and an anode that accelerate electrons up to 180V into neutral gas, ionizing it. The machine is produces plasma discharges at a rate of up to 1 Hz and has a comprehensive set of plasma diagnostics that can be placed virtually anywhere on the machine. The magnetic geometry is controlled by 13 independent power supplies, and is fueled by either a volumetric fill system, gas puffing, or both.

Experiments with 67 different machine configurations (different scalar settings) were performed on the LAPD, totaling around 130 thousand discharges. There were two phases of data collection, \texttt{DR1} and \texttt{DR2}, that had differing machine behavior. Four experimental configurations from each of these two phases were held out as representative configurations composing the test set.

The dataset includes nine time series (of length 76) in total: discharge current, discharge voltage, 5 visible light diode signals (the last of which having a He-II filter), interferometer (line-integrated density), and ion saturation current -- $I_\text{sat} ~ \sim n_e \sqrt{T_e}$ where $n_e$ is electron density and $T_e$ is electron temperature -- from a Langmuir probe. All time-series data were downsampled to a common sampling rate of 2.5 kHz and truncated to match length. Additionally, fifteen scalar inputs and flags were included: probe position (x, y, z), magnetic field (source, mirror, and midplane), gas puff duration and voltage, total gas pressure, first 4 amu RGA readings, run set flag, and top gas puff flag. In total, the input length is 699 features when concatenated. Each input is mean-subtracted and scaled independently by the peak-to-peak value. Scaling each value was necessary because the variance in intensity of the signals varied considerably over the time series and learning distributions of very small width appears to be difficult for EBMs.

\section{Training the EBM}
The following loss function is used to train the EBM:
\begin{equation}
	\mathcal{L} = \mathcal{L}_\text{CD} + \mathcal{L}_\text{KL} + \alpha \mathcal{L}_\text{reg}
\end{equation}
where $\mathcal{L}_\text{CD}$ is the contrastive divergence (CD) loss, $\mathcal{L}_\text{KL}$ is the KL-divergence loss, and $\mathcal{L}_\text{reg}$ is the energy regularization loss, listed in order of importance. The algorithm for training and loss function breakdown can be seen in \ref{alg:training}.

\begin{algorithm}
\caption{EBM training algorithm}\label{alg:training}
\begin{algorithmic}
\Require Training samples $x_i^+$, training data distribution $\mathcal{}p_D$, energy function $E_\theta$, replay buffer $\mathcal{B}$, step size $\epsilon$, MCMC steps $L$, KL MCMC steps $K$, energy regularization strength $\alpha$, stop gradient operator $\Omega(\cdot)$, replay fraction $f_\mathcal{B}$, batch size $M$

	\State $\mathcal{B} \gets \mathcal{U}(-1,1)$ \Comment{Fill buffer from uniform distribution}

	\While{not converged}
		\State $x_i^+ \sim \mathcal{}p_D$
		\State $\tilde{x}_{i}^0 \sim \mathcal{B}$ sample $M f_\mathcal{B}$ negative examples, $\mathcal{U}(-1,1)$ otherwise
		\State $X \sim \mathcal{B}$ nearest-neighbor samples such that $X \cap \tilde{x}_{i}^0 = \varnothing$
		\\
		\For{sample step $\ell=1$ to L}  \Comment{Run Langevin dynamics}
			\State $\tilde{x}_{i}^\ell \gets \tilde{x}_{i}^{\ell-1} - \frac{\epsilon^2}{2}\nabla_x E_\theta(\tilde{x}_{i}^{\ell-1}) + \epsilon \mathcal{N}(0, 1)$
		\EndFor
		\State $\tilde{x}_{i}^{L} = \Omega(\tilde{x}_{i}^{L})$
		\\	
		\State $\hat{x}_{i}^0 = \tilde{x}_{i}^\ell$ where $\ell = L-K$ \Comment{Run Langevin dynamics for KL loss}
		\For{KL sample step $k=1$ to K}
			\State $\hat{x}_{i}^k \gets \hat{x}_{i}^{k-1} - \frac{\epsilon^2}{2}\nabla_x E_\theta(\hat{x}_{i}^{k-1}) + \epsilon \mathcal{N}(0, 1)$ 
		\EndFor
		\\
		\State $\mathcal{L}_\text{CD} = \frac{1}{M} \sum_i E_\theta(\tilde{x}_{i}^+) - E_\theta(\tilde{x}_{i}^L)$ 
		\State $\mathcal{L}_\text{KL} = \frac{1}{M} \sum_i E_{\Omega(\theta)}(E_\theta(\hat{x}_{i}^K) - \text{NN}(X, \hat{x}_{i}^K)$ \Comment{Has gradients through MCMC}
		\State $\mathcal{L}_\text{reg} = \frac{1}{M} \sum_i E_\theta(\tilde{x}_{i}^+)^2 + E_\theta(\tilde{x}_{i}^L)^2$
		
		\State $\mathcal{L} = \mathcal{L}_\text{CD} + \mathcal{L}_\text{KL} + \alpha \mathcal{L}_\text{reg}$
		\\		
		\State Apply $\nabla_\theta \mathcal{L}$ to $\theta$ via the Adam optimizer

		\State $\mathcal{B} \gets \mathcal{B} \cup \mathcal{U}(-1,1)$ and remove samples to maintain buffer size
	\EndWhile
	
\end{algorithmic}
\end{algorithm}

\subsection{The sampler}

The sampler is one of the most important considerations when building the EBM. The samples are initialized from random noise between $-1$ and 1. We use Langevin dynamics to move the samples as formulated in Nijkamp et al. \cite{nijkamp_anatomy_2020}, as seen in algorithm \ref{alg:sampling}. The number of sampler steps per batch was critical for training stability: 30 appears to be close to the minimum number of steps for stable training, similar to what has been observed in other studies \cite{cheng_versatile_2024}. The step size is typically chosen to match the standard deviation of the smallest feature, but that was found to be too large for this use case. Given that our data are highly multimodal, some modes of the input distribution had a deviation of 0 (such as the flags) even after scaling so smaller step size of $\epsilon = 0.01$ was used. 

\begin{algorithm}
\caption{EBM sampling}\label{alg:sampling}
\begin{algorithmic}
\Require Energy function $E_\theta$, auxiliary energy function $F$, step size $\epsilon$, MCMC sampling steps $L$
	\State $\tilde{x}_{i}^0 \sim \mathcal{U}(-1,1)$ \Comment{Initialize on uniform distribution}
	\For{sample step $\ell=1$ to L}  \Comment{Run Langevin dynamics}
		\State $\tilde{x}_{i}^\ell \gets \tilde{x}_{i}^{\ell-1} - \frac{\epsilon^2}{2}\nabla_x \left( E_\theta(\tilde{x}_{i}^{\ell-1}) + F(\tilde{x}_{i}^{\ell-1}) \right)+ \epsilon \mathcal{N}(0, 1)$
	\EndFor
	\\
	\State Return $\tilde{x}_{i}^{L}$
	
\end{algorithmic}
\end{algorithm}

A replay buffer is used to provide a warm start for the sampler chains. Every batch iteration, 5\% of the samples from the buffer are replaced with random noise (a replay fraction of 0.95). 

The model was trained for 27 hours over 172 epochs. The batch size was 128 and was trained with the AdamW optimizer with a learning rate of $10^{-4}$, decaying to $10^{-5}$ after 20 epochs using a step function schedule. The first 100 epochs were trained in a single run; the next 72  were resumed from the final checkpoint of the previous run, leading to reinitialization of the replay buffer. Given the low learning rate, this did not have a significant impact on model training. This model, taken from the 39th epoch of the second run, was used for inference; after which, training diverged. 

\begin{figure}
	\centering
	\includegraphics[width=300pt]{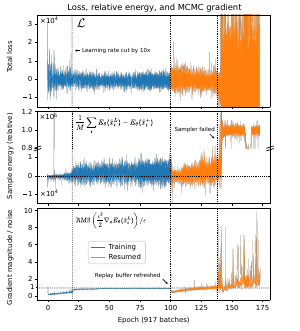}
	\caption[EBM training curves]{\label{fig:losses-energies}Training curves of the model. Top: the total loss, middle: the relative energies of the negative samples and the training examples (positive samples), bottom: the gradient of the energy function normalized to the noise scale (step size $\epsilon$).}
\end{figure}

The training curves of loss, relative energy, and energy gradient can be seen in fig. \ref{fig:losses-energies}. The loss curve is difficult to interpret given the composite nature of the loss function. However, maintaining a relative energy near 0 between negative and positive samples indicates that the generated samples are close to the data distribution. When the relative energies diverge, samples are no longer accurately representing the data and training typically collapses. The energy function learns the scale of the gradients required to produce good samples. 

\section{Architecture}

\begin{figure}
    \centering
    \begin{subfigure}[c]{0.45\textwidth}
        \centering
        \includegraphics[width=\linewidth]{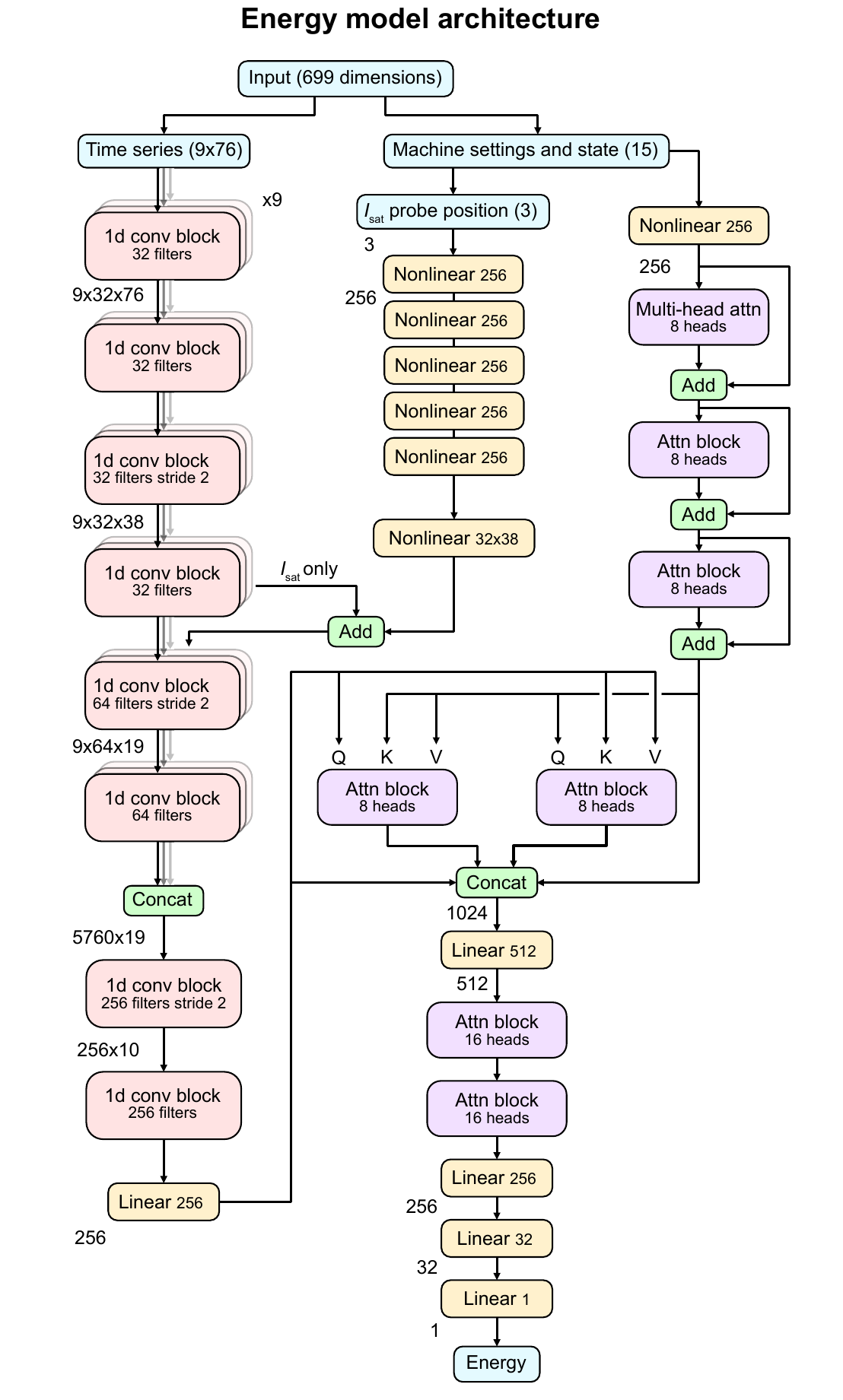}
        \caption[EBM neural network architecture]{\label{fig:architecture}Architecture for learning the energy function. Two main branches were used: processing the time series inputs and the machine settings. The layer blocks are defined in fig. \ref{fig:architecture_blocks}.}
    \end{subfigure}
    \hfill
    \begin{subfigure}[c]{0.4\textwidth}
        \centering
        \includegraphics[width=\linewidth]{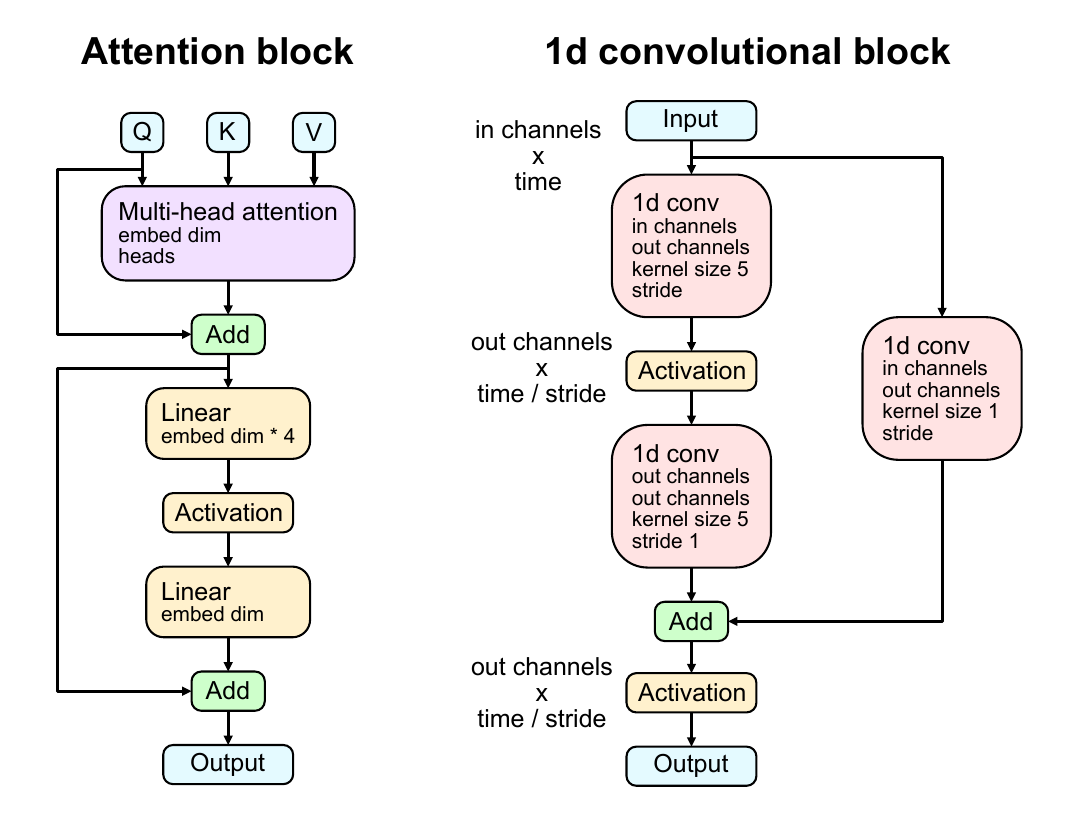}
        \caption{\label{fig:architecture_blocks}Layer blocks used in the EBM architecture. Residual layers work well.}
    \end{subfigure}
\end{figure}

The model is intrinsically multi-modal: time-series data from diagnostics is mixed in with machine settings, state, and probe position. In this model we preprocess separately each time series and the LAPD configuration. Convolutional NNs were used for the time series input, and transformer-like multi-head attention blocks were used for the settings, state, and probe position. The time series convolutions were merged in another convolutional pass, and the two branches were combined using multi-head attention.  A visual representation of the architecture can be seen in fig. \ref{fig:architecture}, with the layer blocks defined in fig. \ref{fig:architecture_blocks}. Convolutions were chosen for the time series input because they are relatively parameter efficient.  The depth of the network guarantees that the receptive field of the downstream neurons covers the entire input space so the positional dependence of the signal is maintained. No pooling layers were used. In general, purely fully-connected networks were found to be difficult to train, consistent with other studies \cite{cheng_versatile_2024}.

Separating signal processing into branches before sensor fusion was necessary for good model performance. The energy layers -- responsible for combining the learned signal representations -- were not as sensitive to parameter count. The EBM struggled to model the positional dependence of the $I_\text{sat}$ signal, so a branch was added from the probe positions to add onto the intermediate representation of the $I_\text{sat}$ signal, improving performance of the positional mapping.

\section{Unconditional sampling}

Unconditional samples can tell us how well the network is modeling the data distribution. 5000 samples were generated with the inputs initialized from a uniform distribution between -1 and 1. These samples ran for 120 steps of Langevin dynamics with the default step size of 0.01. On an RTX 3090, this process took 64 seconds. 

\begin{figure}
	\centering
	\includegraphics[width=\linewidth]{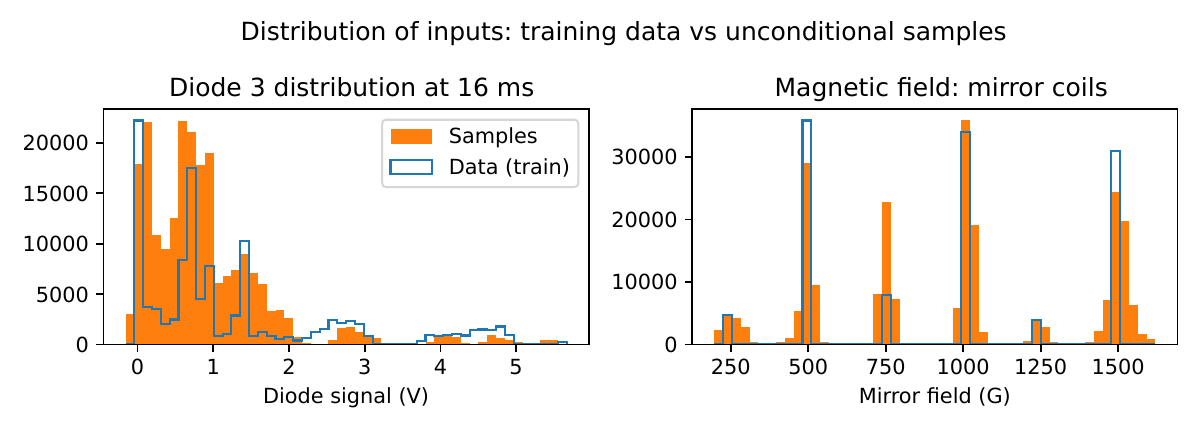}
	\caption[Unconditional samples -- diode 3 and mirror field]{\label{fig:uncond_examples}Unconditional samples of diode 3 at 16 ms and the mirror coil magnetic field inputs, chosen as representative examples. The EBM learns all modes of the distributions, though the probability mass is not well-aligned.}
\end{figure}

Given that the data distributions are highly multi-modal, it is important that the EBM captures many, if not all, modes of the distribution. Representative examples of these distributions, namely of diode 3 at 16 ms (chosen arbitrarily) and the mirror field, can be seen in fig. \ref{fig:uncond_examples}. The density of these particular modes often differs between the samples and the real data, indicating that there is ample room for improved training and development of this model.

\section{Diagnostic reconstruction via conditional sampling}

Conditional sampling of these models is typically performed by freezing the gradients of the inputs to be conditioned on. In this work, we instead modify the energy function based on the data to be conditioned on. Given the energy model $E(x)$, where $x$ is an input into the model, we add an auxiliary energy function $F(x)$ that is added to $E(x)$. This creates a new aggregate energy function $E(x) + F(x)$. By definition of energy $p_E(x) \propto \exp(-E(x))$, this is equivalent to multiplying the two distributions $p_E(x) \cdot p_F(x)$. In other words, adding this auxiliary energy function $F(x)$ to $E(x)$ implies we are sampling over the distribution $p_F \cap p_F$.

The choice of auxiliary energy function $F(x)$ can make a large difference on the conditional samples. We choose $F(x)$ to be a quadratic energy function centered on the data: 
\begin{equation}
	F(x) = \left(\frac{x - x_\text{fixed}}{\sigma_F} \right)^2
\end{equation}
where $\sigma_F$ controls the horizontal scale of the quadratic function. Interpreted as a probability distribution, this is a Gaussian with standard deviation $\sigma_F$. The minimal width for stable sampling appears to be $\approx (2 \epsilon)^2$ which matches intuition: if the width of $F(x)$ approaches the step size, a Langevin update step may place the particle at a much higher energy with much larger gradients.
 This behavior ties the conditional sampling directly to the training process: the fidelity of the model is set by the step size.
 Other distributions were used, such as a Laplace distribution via $F(x) = \vert x - x_\text{fixed} \vert$, but the samples produced lacked the diversity seen when using a quadratic $F(x)$. Note that the normalizing constants of the probability distribution $p_F(x)$ do not matter because they vanish after $-\log(\cdot)$ is applied and the energy gradients $\nabla_x F(x)$ are taken.

\subsection{Sampling interferometer signals}

\begin{figure}
	\centering
	\includegraphics[width=\linewidth]{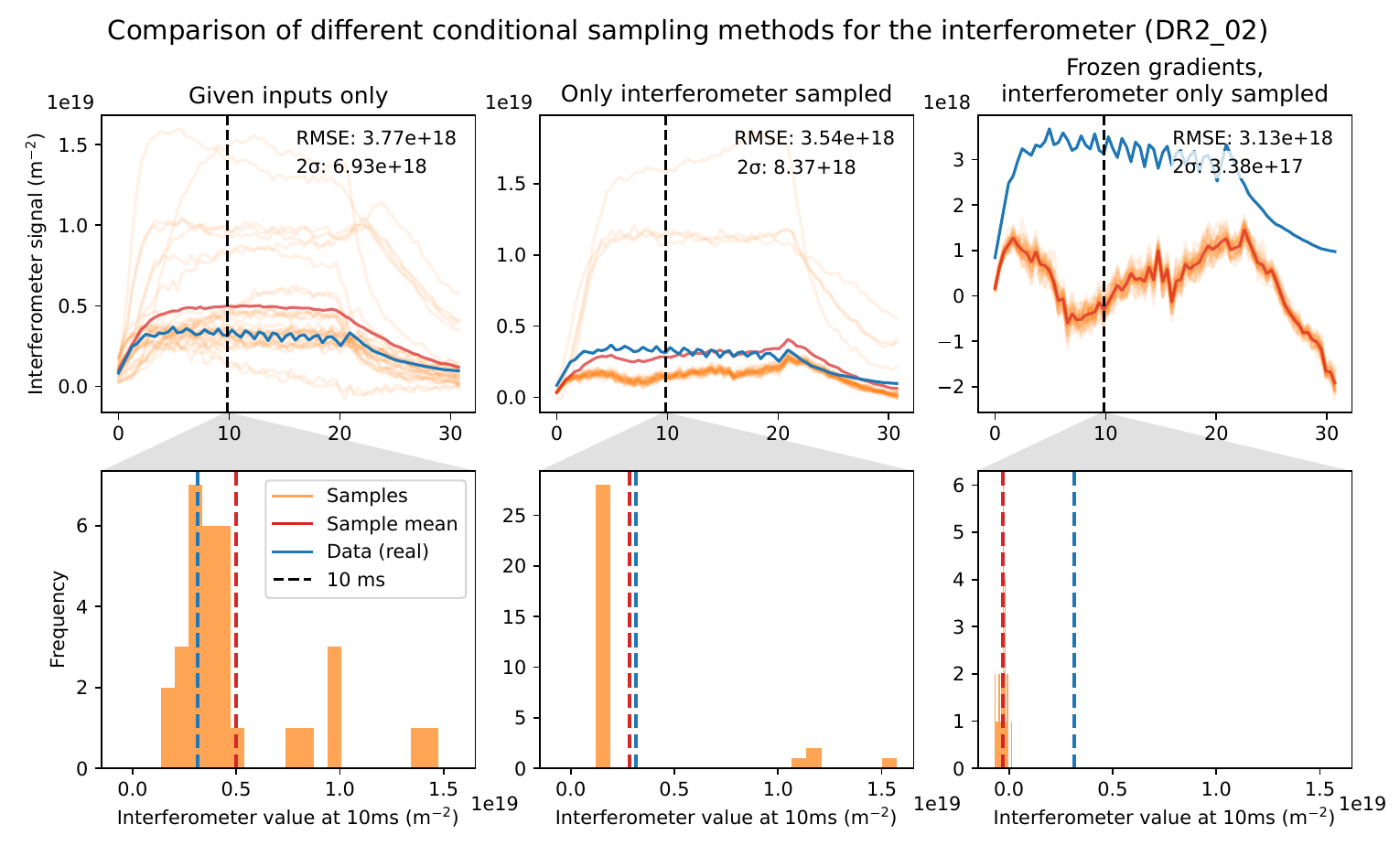}
	\caption[Interferometer time series reconstruction]{\label{fig:ifo_sample}Reconstructing the interferometer signal for a test-set datarun, showing only 32 samples for clarity. Given only the inputs (left), the interferometer signal reconstruction uncertainty is quite large with many possible modes. When given other diagnostics signals, the RMSE improves by $2 \times 10^{17}$ m$^{-2}$, but the uncertainty increases. If the model is sampled by instead initializing all inputs on real data and freezing the gradients (right), the model produces unphysical results and is poorly calibrated. The datarun chosen (DR2\_02) is representative of performance across the test set. }
\end{figure}

Here, we choose to reconstruct interferometer signals because the results are more easily interpreted physically than the diodes and the discharge current. Using conditional sampling, we compare the samples when only the LAPD control inputs are given and compare with sampling when other diagnostics are given -- namely the discharge current, diodes, and $I_\text{sat}$. The machine inputs are the discharge voltage, gas puff duration and voltage, gas pressures, and magnetic field configuration. We also compare the traditional method of initializing on data by freezing gradients. A summary of the results and standard deviation of the distributions can be seen in table. \ref{tab:ifo-cond-sample}. The full time series of the samples can be seen in fig. \ref{fig:ifo_sample}. 

\begin{table*}
\small
	\centering
	\caption{RMSE and 2$\sigma$ of conditional interferometer samples for test set and \texttt{DR2\_02}}
	\label{tab:ifo-cond-sample}
	\begin{tabular}{l l l l}
		Given: & LAPD settings only & All signals & Frozen gradients \\
		\hline
		RMSE (test set) & $4.12 \times 10^{18}$ & $2.91 \times 10^{18}$ & $3.13 \times 10^{18}$ \\
		RMSE (\texttt{DR2\_02})& $3.77 \times 10^{18}$ & $3.54 \times 10^{18}$ & $2.51 \times 10^{18}$ \\
		2 $\sigma$ (\texttt{DR2\_02}) & $6.93 \times 10^{18}$ & $8.37 \times 10^{18}$ & $3.38 \times 10^{17}$ \\
		\hline
		Training RMSE & $4.40 \times 10^{17}$ \\
	\end{tabular}
\end{table*}

For sampling, 90 samples steps were taken with the training step size of $\epsilon=0.01$. The auxiliary energy function used a width of $(2\epsilon)^2$. Interferometer traces from a single shot from each of the eight test set dataruns were sampled. 128 samples of the interferometer signal were taken from each datarun, taking approximately 9 seconds on an RTX 3090. 

Sampling with only machine inputs given led to a large variety of potential interferometer signals, seen on the left of fig. \ref{fig:ifo_sample}. When providing additional signals, the variety of the sample distribution is dramatically decreased (seen in the bottom row), the shapes of the curves when diagnostics are given matches better, and the RMSE improves as seen in tab. \ref{tab:ifo-cond-sample}. The $2 \sigma$ range of the samples increases, however, perhaps indicating increased uncertainty. 

Sampling while freezing gradients (right side of fig. \ref{fig:ifo_sample}) can lead to good RMSE values relative to the other samples, but the actual samples are unphysical: negative density is impossible. Constraining the samples to be greater than 0 using an auxiliary energy function ensures positive interferometer values, but the sample quality remains very poor and increases the RMSE. The distribution of signals is also very narrow, leading to a very overconfident prediction. 

Any diagnostic in the dataset can be conditionally sampled given other data, not only the interferometer; this is a powerful feature which may enable opportunities such as post hoc diagnostic calibration.

\section{Trend inference: visible light as a function of discharge power}

This learned EBM can also be used to infer trends in discharge behavior. We are concerned with the relative amount of plasma indicated by visible light emitted at the end of the machine furthest from the plasma-generating cathode. This light is measured by diode \#3. We seek to understand the relationship of this visible light measurement to discharge power $P= VI$.

\begin{figure}
	\centering
	\includegraphics[width=250pt]{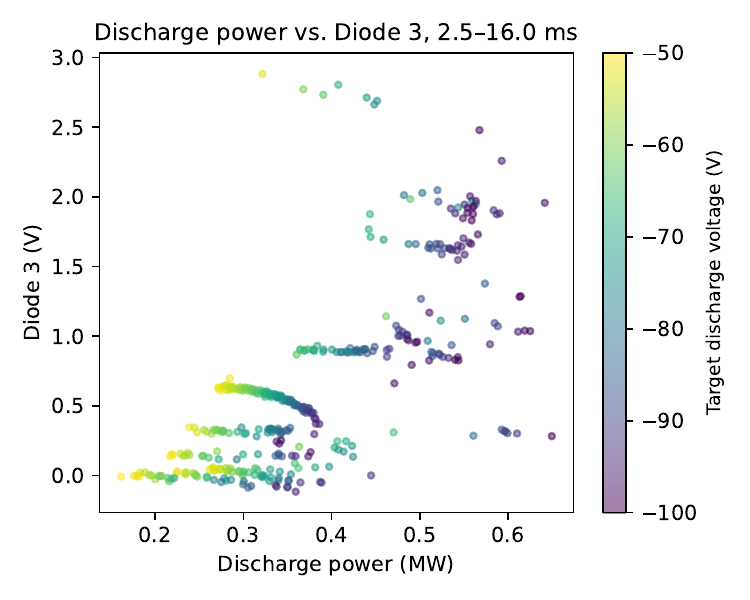}
	\caption[Diode \#3 signal as a function of discharge power]{\label{fig:power_diode_scan}Diode \#3 signal as a function of discharge power. As power increases, the amount of visible light at the end of the machine also increases. Note that increased discharge voltage does not always lead to higher discharge power.}
\end{figure}

Using the conditional sampling method outlined in the previous section, we scan over a set discharge voltages averaged between 2.5 and 16 ms (avoiding sharp transients) between -50V and -100V in 1V increments. The magnetic field was  constrained to a uniform 1 kG, the probe positions to (0, 0, 600) cm, the gas puff duration to 38 ms, and the top gas puff valve off. The discharge voltage auxiliary energy function was set to a width of $(2\epsilon)^2$, but it was found that the machine configuration auxiliary energy function worked better around a width $(1\epsilon)^2$. The gas puff pressures and valve settings were left to freely vary. 8 samples were taken at each discharge value, each sampled for 90 steps over the energy surface. The relationship between the discharge power and diode \#3 signal can be seen in fig. \ref{fig:power_diode_scan}. 

In general, the discharge power leads to higher visible light, and thus inferred plasma, at the end of the LAPD furthest from the cathode. Thus, if we wish to have a plasma with the maximum column length, we should design our discharges for maximum power. Notably, the discharge voltage does not always lead to higher discharge power: the discharge current is affected by the downstream condition and configuration of the LAPD. There are clear limitations in the coverage of machine state space in the dataset collected, as seen in the striations in the scatterplot. A densely-filled dataset would not lead to these spurious valleys in the energy function. Nevertheless, we demonstrate that this EBM can be used to infer trends and gain understanding under uncertainty in machine configuration (i.e., gas pressure and fueling) which would be difficult, if not impossible, to do so experimentally.

\section{Insight via symmetries in the energy function}

\begin{figure}
	\centering
	\includegraphics[width=300pt]{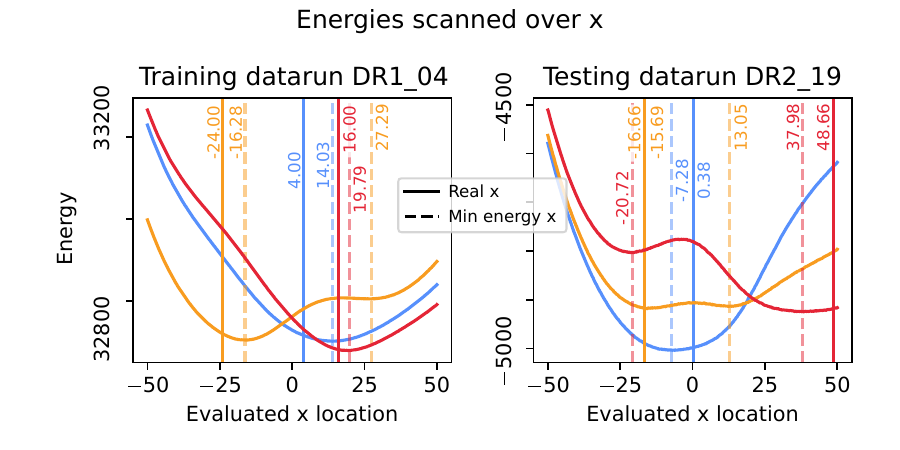}
	\caption[Energy function scan along probe x coordinate]{\label{fig:energy_x_scan}Scans along the x-axis input for the energy function of a real shot. When off-axis shots are provided, the energy function may encode the symmetry about the y axis.}
\end{figure}

Structure in the energy function itself can also be examined to extract insights from the data. One example of this can be seen in scans along the energy function for the x-axis probe position. 
Given a certain $I_\text{sat}$ time series, we expect the position of the probe to be equally likely at a constant radius if the azimuthal symmetry is perfect given the cylindrical geometry of the LAPD. Energy scans along the x axis for a given shot (and $I_\text{sat}$ time series) for a training and a test set datarun can be seen in fig. \ref{fig:energy_x_scan}. When a shot near or on the magnetic axis is provided, the energy function is largely symmetric about that point. When a shot is provided further out, the energy function takes on a shape with two minima, indicating that two positions of the energy surface are likely. This behavior is obvious in the test set case of \texttt{DR2\_19} (yellow curve) -- either side of the machine center nearly equally likely. For a shot near the machine wall (red curve), the energy distribution has two minima: the minimum on the incorrect side is closer than ideal, but the true symmetrical position is beyond the limits of the training data. Not all shots yield symmetric energy functions, as seen in the red curve of the training set (left side of fig. \ref{fig:energy_x_scan}). 

This direct analysis of the energy function is uncommon, if not novel, owing to the intrinsic structure of the data. Analyzing energy functions in this way requires a dataset that has strong dependence on a low dimensional subset -- sufficiently low that a comprehensive grid scan is tractable. The probe coordinates and machine settings satisfy this condition: changing one of those values dramatically affects likely values of other inputs. 

This direct analysis of the energy function highlights the ability of examining solutions to ill-posed inverse problems which would be difficult with typical regression techniques.

\section{Summary, limitations, and future work}

In this work, we use a multi-branch, medium-term fusion architecture, signals distribution over a highly multimodal dataset in both definitions of the term: the distribution has many modes and the inputs are different modalities. Using this single EBM, we reconstruct missing diagnostic signals, uncover symmetries in probe position, and infer trends in discharge power. We accomplish these tasks by directly evaluating the energy function or conditionally sampling from the EBM by utilizing an auxiliary energy function. 

This work is fundamentally limited by the lack of diversity in the dataset and lack of exploration in model architecture and training hyperparameters. Improved dataset diversity would alleviate potentially spurious correlations seen in conditional samples, and would improve confidence in the trends inferred from the model. In addition, LAPD plasma parameters drift over time without changing configuration, limiting the insight gained to the time period the data were collected. The architecture also has ample room for improvement as development of the architecture stopped as soon it was good enough. In addition, much uncertainty remains in how to effectively choose the EBM hyperparameters, particularly in the interaction of the regularization term, energy surface gradient magnitude, and the capacity of the network. 

Nevertheless, results are promising and encourage future work. A 29-million LAPD discharge dataset is available for training, which could greatly improve the predictive capabilities of the model as the LAPD evolves. In addition, the flexibility of EBMs permits joint sampling with learned theoretical models, potentially providing a physical constraint on possible plasma configurations, or vice versa. Regardless, generative modeling will likely become an increasingly important tool in plasma data analysis.

\section{Acknowledgements}

The authors would like to thank Professor Christoph Niemann, Dr. Shreekrishna Tripathi, Dr. Steve Vincena, Dr. Pat Pribyl Tom Look, Yhoshua Wug, Dr. Lukas Rovige, and Dr. Jia Han for experimental support. This work was performed at the Basic Plasma Science Facility, which is a DOE Office of Science, FES collaborative user facility, and is funded by DOE (DE-FC02-07ER54918)

\bibliographystyle{plainnat}
\bibliography{neurips_2026.bib}

@article{LAPD_LaB6,
    author = {Qian, Yuchen and Gekelman, Walter and Pribyl, Patrick and Sketchley, Tom and Tripathi, Shreekrishna and Lucky, Zoltan and Drandell, Marvin and Vincena, Stephen and Look, Thomas and Travis, Phil and Carter, Troy and Wan, Gary and Cattelan, Mattia and Sabiston, Graeme and Ottaviano, Angelica and Wirz, Richard},
    title = "{Design of the Lanthanum hexaboride based plasma source for the large plasma device at UCLA}",
    journal = {Review of Scientific Instruments},
    volume = {94},
    number = {8},
    pages = {085104},
    year = {2023},
    month = {08},
    issn = {0034-6748},
    doi = {10.1063/5.0152216},
    url = {https://doi.org/10.1063/5.0152216},
    eprint = {https://pubs.aip.org/aip/rsi/article-pdf/doi/10.1063/5.0152216/18074441/085104\_1\_5.0152216.pdf},
}

@article{travis_machine-learned_2025,
	title = {Machine-learned trends in mirror configurations in the large plasma device},
	volume = {32},
	issn = {1070-664X, 1089-7674},
	url = {https://pubs.aip.org/pop/article/32/8/082106/3357758/Machine-learned-trends-in-mirror-configurations-in},
	doi = {10.1063/5.0270755},
	number = {8},
	urldate = {2025-11-12},
	journal = {Physics of Plasmas},
	author = {Travis, Phil and Bortnik, Jacob and Carter, Troy},
	month = aug,
	year = {2025},
	pages = {082106},
}

@misc{cheng_versatile_2024,
	title = {Versatile {Energy}-{Based} {Probabilistic} {Models} for {High} {Energy} {Physics}},
	url = {http://arxiv.org/abs/2302.00695},
	doi = {10.48550/arXiv.2302.00695},
	urldate = {2025-04-18},
	publisher = {arXiv},
	author = {Cheng, Taoli and Courville, Aaron},
	month = jan,
	year = {2024},
	note = {arXiv:2302.00695 [cs]},
}

@phdthesis{lennart_van_rijn_minimizing_2022,
	title = {Minimizing neoclassical transport in the {Wendelstein} 7-{X} stellarator using variational autoencoders},
	school = {Eindhoven University of Technology},
	author = {{Lennart van Rijn}},
	month = jul,
	year = {2022},
}

@phdthesis{vos_discovery_2024,
	title = {Discovery of hidden {Neoclassical} {Transport} variables in {Wendelstein} 7-{X} through {Variational} {AutoEncoder} {Latent} {Space} {Exploration}},
	author = {Vos, J M},
	year = {2024},
}

@article{dave_synthetic_2023,
	title = {Synthetic data generation using generative adversarial network for tokamak plasma current quench experiments},
	volume = {63},
	issn = {0863-1042, 1521-3986},
	url = {https://onlinelibrary.wiley.com/doi/10.1002/ctpp.202200051},
	doi = {10.1002/ctpp.202200051},
	number = {5-6},
	urldate = {2025-04-23},
	journal = {Contributions to Plasma Physics},
	author = {Dave, Bhrugu and Patel, Sarthak and Shivani, Rishi and Purohit, Shishir and Chaudhury, Bhaskar},
	month = jun,
	year = {2023},
	pages = {e202200051},
}

@article{daly_data-driven_2023,
	title = {Data-driven plasma modelling: surrogate collisional radiative models of fluorocarbon plasmas from deep generative autoencoders},
	volume = {4},
	issn = {2632-2153},
	shorttitle = {Data-driven plasma modelling},
	url = {https://iopscience.iop.org/article/10.1088/2632-2153/aced7f},
	doi = {10.1088/2632-2153/aced7f},
	number = {3},
	urldate = {2025-04-23},
	journal = {Machine Learning: Science and Technology},
	author = {Daly, G A and Fieldsend, J E and Hassall, G and Tabor, G R},
	month = sep,
	year = {2023},
	pages = {035035},
}

@article{skvara_detection_2020,
	title = {Detection of {Alfvén} {Eigenmodes} on {COMPASS} with {Generative} {Neural} {Networks}},
	volume = {76},
	issn = {1536-1055, 1943-7641},
	url = {https://www.tandfonline.com/doi/full/10.1080/15361055.2020.1820805},
	doi = {10.1080/15361055.2020.1820805},
	number = {8},
	urldate = {2025-04-23},
	journal = {Fusion Science and Technology},
	author = {Škvára, Vít and Šmídl, Václav and Pevný, Tomáš and Seidl, Jakub and Havránek, Aleš and Tskhakaya, David},
	month = nov,
	year = {2020},
	pages = {962--971},
}

@inproceedings{juven_generative_2024,
	address = {London, United Kingdom},
	title = {Generative {Models} and {Simulation} to {Assess} {Uncertainties} for {Tokamak} {Infrared} {Thermography}},
	copyright = {https://doi.org/10.15223/policy-029},
	isbn = {9798350372250},
	url = {https://ieeexplore.ieee.org/document/10734728/},
	doi = {10.1109/MLSP58920.2024.10734728},
	urldate = {2025-04-23},
	booktitle = {2024 {IEEE} 34th {International} {Workshop} on {Machine} {Learning} for {Signal} {Processing} ({MLSP})},
	publisher = {IEEE},
	author = {Juven, Alexis and Aumeunier, Marie-Hélène and Marot, Julien},
	month = sep,
	year = {2024},
	pages = {1--6},
}

@article{pavone_machine_2023,
	title = {Machine learning and {Bayesian} inference in nuclear fusion research: an overview},
	volume = {65},
	issn = {0741-3335, 1361-6587},
	shorttitle = {Machine learning and {Bayesian} inference in nuclear fusion research},
	url = {https://iopscience.iop.org/article/10.1088/1361-6587/acc60f},
	doi = {10.1088/1361-6587/acc60f},
	number = {5},
	urldate = {2025-04-23},
	journal = {Plasma Physics and Controlled Fusion},
	author = {Pavone, A and Merlo, A and Kwak, S and Svensson, J},
	month = may,
	year = {2023},
	pages = {053001},
}

@misc{jalalvand_multimodal_2024,
	title = {Multimodal {Super}-{Resolution}: {Discovering} hidden physics and its application to fusion plasmas},
	shorttitle = {Multimodal {Super}-{Resolution}},
	url = {http://arxiv.org/abs/2405.05908},
	doi = {10.48550/arXiv.2405.05908},
	urldate = {2025-05-05},
	publisher = {arXiv},
	author = {Jalalvand, Azarakhsh and Kim, SangKyeun and Seo, Jaemin and Hu, Qiming and Curie, Max and Steiner, Peter and Nelson, Andrew Oakleigh and Na, Yong-Su and Kolemen, Egemen},
	month = nov,
	year = {2024},
	note = {arXiv:2405.05908 [physics]},
}

@article{wei_dimensionality_2021,
	title = {A dimensionality reduction algorithm for mapping tokamak operational regimes using a variational autoencoder ({VAE}) neural network},
	volume = {61},
	issn = {0029-5515, 1741-4326},
	url = {https://iopscience.iop.org/article/10.1088/1741-4326/ac3296},
	doi = {10.1088/1741-4326/ac3296},
	number = {12},
	urldate = {2025-04-24},
	journal = {Nuclear Fusion},
	author = {Wei, Y. and Levesque, J.P. and Hansen, C.J. and Mauel, M.E. and Navratil, G.A.},
	month = dec,
	year = {2021},
	pages = {126063},
}

@article{du_improved_2021,
	title = {Improved {Contrastive} {Divergence} {Training} of {Energy} {Based} {Models}},
	url = {http://arxiv.org/abs/2012.01316},
	urldate = {2021-11-21},
	journal = {arXiv:2012.01316 [cs]},
	author = {Du, Yilun and Li, Shuang and Tenenbaum, Joshua and Mordatch, Igor},
	month = jun,
	year = {2021},
	note = {arXiv: 2012.01316}
}

@article{du_implicit_2020,
	title = {Implicit {Generation} and {Generalization} in {Energy}-{Based} {Models}},
	url = {http://arxiv.org/abs/1903.08689},
	urldate = {2021-02-23},
	journal = {arXiv:1903.08689 [cs, stat]},
	author = {Du, Yilun and Mordatch, Igor},
	month = jun,
	year = {2020},
	note = {arXiv: 1903.08689},
}

@misc{du_compositional_2020,
	title = {Compositional {Visual} {Generation} and {Inference} with {Energy} {Based} {Models}},
	url = {http://arxiv.org/abs/2004.06030},
	doi = {10.48550/arXiv.2004.06030},
	urldate = {2025-04-18},
	publisher = {arXiv},
	author = {Du, Yilun and Li, Shuang and Mordatch, Igor},
	month = dec,
	year = {2020},
	note = {arXiv:2004.06030 [cs]},
}

@misc{du_unsupervised_2021,
	title = {Unsupervised {Learning} of {Compositional} {Energy} {Concepts}},
	url = {http://arxiv.org/abs/2111.03042},
	doi = {10.48550/arXiv.2111.03042},
	urldate = {2025-04-18},
	publisher = {arXiv},
	author = {Du, Yilun and Li, Shuang and Sharma, Yash and Tenenbaum, Joshua B. and Mordatch, Igor},
	month = nov,
	year = {2021},
	note = {arXiv:2111.03042 [cs]},
}

@article{lecun_tutorial_2006,
	title = {A {Tutorial} on {Energy}-{Based} {Learning}},
	author = {LeCun, Yann and Chopra, Sumit and Hadsell, Raia and Ranzato, Marc’Aurelio and Huang, Fu Jie},
	year = {2006},
	pages = {59},
}

@misc{nijkamp_learning_2019,
	title = {Learning {Non}-{Convergent} {Non}-{Persistent} {Short}-{Run} {MCMC} {Toward} {Energy}-{Based} {Model}},
	url = {http://arxiv.org/abs/1904.09770},
	urldate = {2022-12-31},
	publisher = {arXiv},
	author = {Nijkamp, Erik and Hill, Mitch and Zhu, Song-Chun and Wu, Ying Nian},
	month = nov,
	year = {2019},
	note = {arXiv:1904.09770 [cs, stat]},
}

@article{nijkamp_anatomy_2020,
	title = {On the {Anatomy} of {MCMC}-{Based} {Maximum} {Likelihood} {Learning} of {Energy}-{Based} {Models}},
	volume = {34},
	issn = {2374-3468, 2159-5399},
	url = {https://aaai.org/ojs/index.php/AAAI/article/view/5973},
	doi = {10.1609/aaai.v34i04.5973},
	number = {04},
	urldate = {2022-08-11},
	journal = {Proceedings of the AAAI Conference on Artificial Intelligence},
	author = {Nijkamp, Erik and Hill, Mitch and Han, Tian and Zhu, Song-Chun and Wu, Ying Nian},
	month = apr,
	year = {2020},
	pages = {5272--5280},
}

@article{bond-taylor_deep_2021,
	title = {Deep {Generative} {Modelling}: {A} {Comparative} {Review} of {VAEs}, {GANs}, {Normalizing} {Flows}, {Energy}-{Based} and {Autoregressive} {Models}},
	shorttitle = {Deep {Generative} {Modelling}},
	url = {http://arxiv.org/abs/2103.04922},
	urldate = {2021-03-30},
	journal = {arXiv:2103.04922 [cs, stat]},
	author = {Bond-Taylor, Sam and Leach, Adam and Long, Yang and Willcocks, Chris G.},
	month = mar,
	year = {2021},
	note = {arXiv: 2103.04922},
}

@article{du_model_2019,
	title = {Model {Based} {Planning} with {Energy} {Based} {Models}},
	url = {https://arxiv.org/abs/1909.06878},
	doi = {https://doi.org/10.48550/arXiv.1909.06878},
	author = {Du, Yilun and Lin, Toru and Mordatch, Igor},
	year = {2019},
	pages = {10},
}

@article{hinton_training_2002,
	title = {Training {Products} of {Experts} by {Minimizing} {Contrastive} {Divergence}},
	volume = {14},
	issn = {0899-7667, 1530-888X},
	url = {https://direct.mit.edu/neco/article/14/8/1771-1800/6687},
	doi = {10.1162/089976602760128018},
	number = {8},
	urldate = {2025-05-15},
	journal = {Neural Computation},
	author = {Hinton, Geoffrey E.},
	month = aug,
	year = {2002},
	pages = {1771--1800},
}

@misc{kingma_auto-encoding_2022,
	title = {Auto-{Encoding} {Variational} {Bayes}},
	url = {http://arxiv.org/abs/1312.6114},
	doi = {10.48550/arXiv.1312.6114},
	urldate = {2025-05-16},
	publisher = {arXiv},
	author = {Kingma, Diederik P. and Welling, Max},
	month = dec,
	year = {2013},
	note = {arXiv:1312.6114 [stat]},
}

@misc{goodfellow_generative_2014,
	title = {Generative {Adversarial} {Networks}},
	url = {http://arxiv.org/abs/1406.2661},
	doi = {10.48550/arXiv.1406.2661},
	urldate = {2025-05-16},
	publisher = {arXiv},
	author = {Goodfellow, Ian J. and Pouget-Abadie, Jean and Mirza, Mehdi and Xu, Bing and Warde-Farley, David and Ozair, Sherjil and Courville, Aaron and Bengio, Yoshua},
	month = jun,
	year = {2014},
	note = {arXiv:1406.2661 [stat]},
}

@misc{carbone_hitchhikers_2024,
	title = {Hitchhiker's guide on {Energy}-{Based} {Models}: a comprehensive review on the relation with other generative models, sampling and statistical physics},
	shorttitle = {Hitchhiker's guide on {Energy}-{Based} {Models}},
	url = {http://arxiv.org/abs/2406.13661},
	doi = {10.48550/arXiv.2406.13661},
	urldate = {2025-04-03},
	publisher = {arXiv},
	author = {Carbone, Davide},
	month = jun,
	year = {2024},
	note = {arXiv:2406.13661 [cs]},
}

@inproceedings{gao_learning_2018,
	address = {Salt Lake City, UT, USA},
	title = {Learning {Generative} {ConvNets} via {Multi}-grid {Modeling} and {Sampling}},
	isbn = {978-1-5386-6420-9},
	url = {https://ieeexplore.ieee.org/document/8579052/},
	doi = {10.1109/CVPR.2018.00954},
	urldate = {2025-05-16},
	booktitle = {2018 {IEEE}/{CVF} {Conference} on {Computer} {Vision} and {Pattern} {Recognition}},
	publisher = {IEEE},
	author = {Gao, Ruiqi and Lu, Yang and Zhou, Junpei and Zhu, Song-Chun and Wu, Ying Nian},
	month = jun,
	year = {2018},
	pages = {9155--9164},
}

@misc{deng_residual_2020,
	title = {Residual {Energy}-{Based} {Models} for {Text} {Generation}},
	url = {http://arxiv.org/abs/2004.11714},
	doi = {10.48550/arXiv.2004.11714},
	urldate = {2025-05-16},
	publisher = {arXiv},
	author = {Deng, Yuntian and Bakhtin, Anton and Ott, Myle and Szlam, Arthur and Ranzato, Marc'Aurelio},
	month = apr,
	year = {2020},
	note = {arXiv:2004.11714 [cs]},
}

@article{ackley_learning_1985,
	title = {A learning algorithm for boltzmann machines},
	volume = {9},
	copyright = {http://doi.wiley.com/10.1002/tdm\_license\_1},
	issn = {03640213},
	url = {https://www.sciencedirect.com/science/article/pii/S0364021385800124},
	doi = {10.1016/S0364-0213(85)80012-4},
	number = {1},
	urldate = {2025-05-16},
	journal = {Cognitive Science},
	author = {Ackley, D and Hinton, G and Sejnowski, T},
	month = mar,
	year = {1985},
	pages = {147--169},
}

@InProceedings{ruslan_deep_2009,
  title = 	 {Deep Boltzmann Machines},
  author = 	 {Salakhutdinov, Ruslan and Hinton, Geoffrey},
  booktitle = 	 {Proceedings of the Twelfth International Conference on Artificial Intelligence and Statistics},
  pages = 	 {448--455},
  year = 	 {2009},
  editor = 	 {van Dyk, David and Welling, Max},
  volume = 	 {5},
  series = 	 {Proceedings of Machine Learning Research},
  address = 	 {Hilton Clearwater Beach Resort, Clearwater Beach, Florida USA},
  month = 	 {16--18 Apr},
  publisher =    {PMLR},
  url = 	 {https://proceedings.mlr.press/v5/salakhutdinov09a.html},
}

@inproceedings{tieleman_training_2008,
	address = {Helsinki, Finland},
	title = {Training restricted {Boltzmann} machines using approximations to the likelihood gradient},
	isbn = {978-1-60558-205-4},
	url = {http://portal.acm.org/citation.cfm?doid=1390156.1390290},
	doi = {10.1145/1390156.1390290},
	urldate = {2025-05-16},
	booktitle = {Proceedings of the 25th international conference on {Machine} learning - {ICML} '08},
	publisher = {ACM Press},
	author = {Tieleman, Tijmen},
	year = {2008},
	pages = {1064--1071},
}

@article{hopfield_neural_1982,
	title = {Neural networks and physical systems with emergent collective computational abilities.},
	volume = {79},
	issn = {0027-8424, 1091-6490},
	url = {https://pnas.org/doi/full/10.1073/pnas.79.8.2554},
	doi = {10.1073/pnas.79.8.2554},
	number = {8},
	urldate = {2025-05-16},
	journal = {Proceedings of the National Academy of Sciences},
	author = {Hopfield, J J},
	month = apr,
	year = {1982},
	pages = {2554--2558},
}

\appendix

\section{Loss function breakdown}
The contrastive divergence loss is defined as:
\begin{equation}
	\mathcal{L}_\text{CD} = \frac{1}{M} \sum_i E_\theta(\tilde{x}_{i}^+) - E_\theta(\tilde{x}_{i}^L)
\end{equation}
This loss calculates the difference in energies between sample from the energy surface $\tilde{x}_{i}^L$ via Langevin dynamics -- the ``negative'' samples -- and the samples from the data distribution $\tilde{x}_{i}^+$, i.e., the training data. This loss places the energy surface in tension, with the surface being pulled down by the data and pushed up on the negative samples. Both terms must be included: absence of negative samples would lead to a vacuous solution such as a flat energy surface.
The KL-divergence loss is:
\begin{equation}
	\mathcal{L}_\text{KL} = \frac{1}{M} \sum_i E_{\Omega(\theta)}(E_\theta(\hat{x}_{i}^K) - \text{NN}(X, \hat{x}_{i}^K)
\end{equation}
This loss, suggested in \cite{du_improved_2021} includes portions of the CD loss typically left out. Specifically, this loss minimizes the sampler energy by propagating gradients through the MCMC steps themselves instead of the final samples. In addition, the nearest-neighbor (\texttt{NN}) samples are used to estimate the energy of the negative samples so that the entropy of the distribution is maximized. This loss, though not necessary, significantly improves training stability. 
The regularization loss
\begin{equation}
	\mathcal{L}_\text{reg} = \frac{1}{M} \sum_i E_\theta(\tilde{x}_{i}^+)^2 + E_\theta(\tilde{x}_{i}^L)^2
\end{equation}
keeps the energy values centered near 0. The absolute value of the energy does not matter -- only the gradients and scale -- but this  is included to keep the results easily interpretable and to avoid extreme energy values as to not run into floating-point representation boundaries. In this work, the scale of this regularization is very small with $\alpha = 1 \times 10^{-6}$. This loss also functions as a great diagnostic for when the sampler fails: the regularization loss will reach $1/\alpha$. The EBM training process is presented in algorithm \ref{alg:training}.

It appears that using the AdamW optimizer with a learning rate scheduler is necessary for training in a reasonable amount of time. Training using stochastic gradient descent required very large learning rates ($10^{-2}$) and could not generate realistic samples even with a learning rate decay schedule. 

\section{Unconditional samples: distribution and energies}
\begin{figure}
	\centering
	\includegraphics[width=300pt]{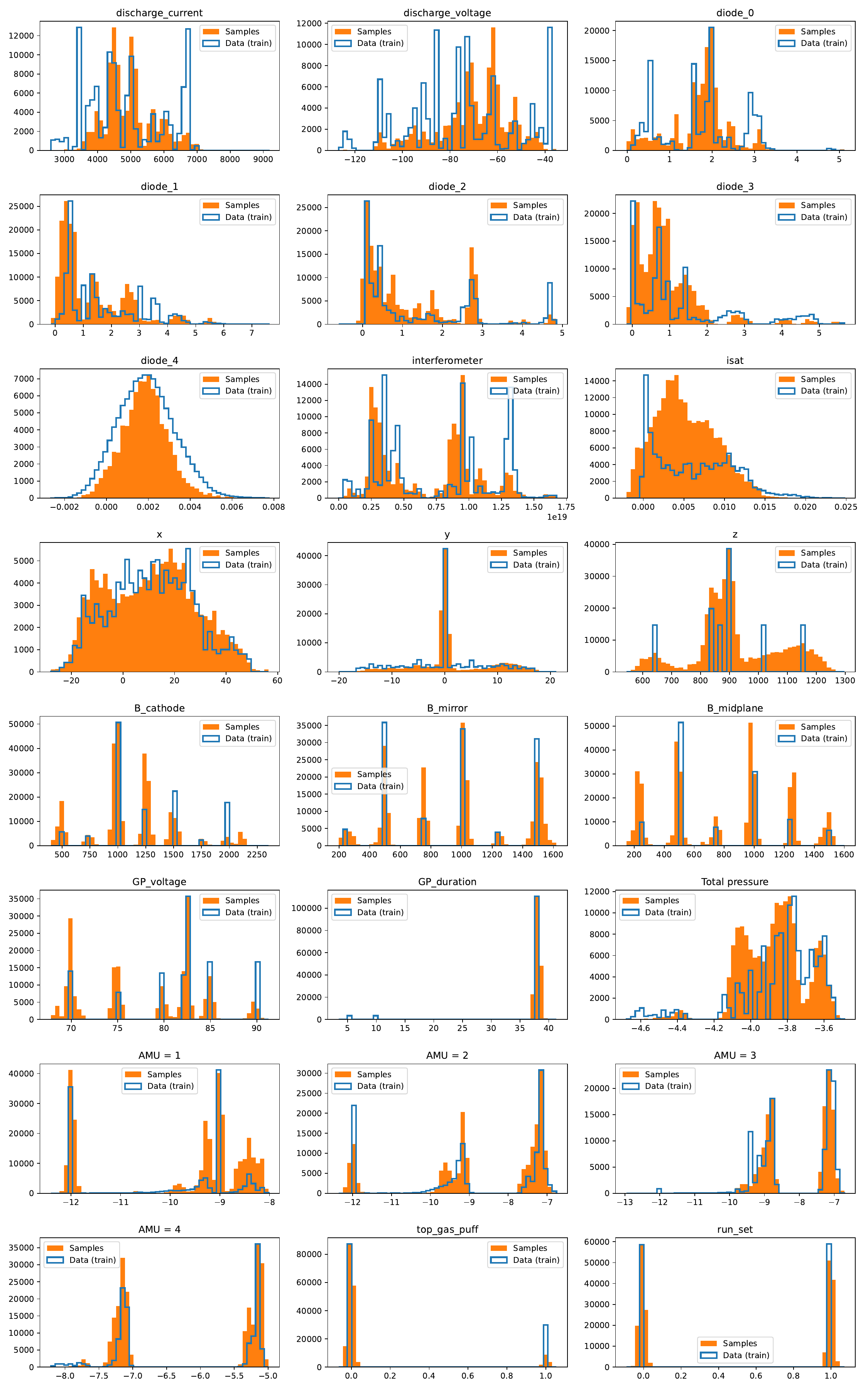}
	\caption[Unconditional samples -- all inputs]{\label{fig:uncond_dist_full}Unconditional samples of all inputs, or at 16 ms -- chosen arbitrarily -- for time series. The model appears to learn most, if not all, modes of the distributions, but can perform poorly when modeling the probability mass, such as in the $I_\text{sat}$ distribution.}
\end{figure}

\begin{figure}
	\centering
	\includegraphics[width=\linewidth]{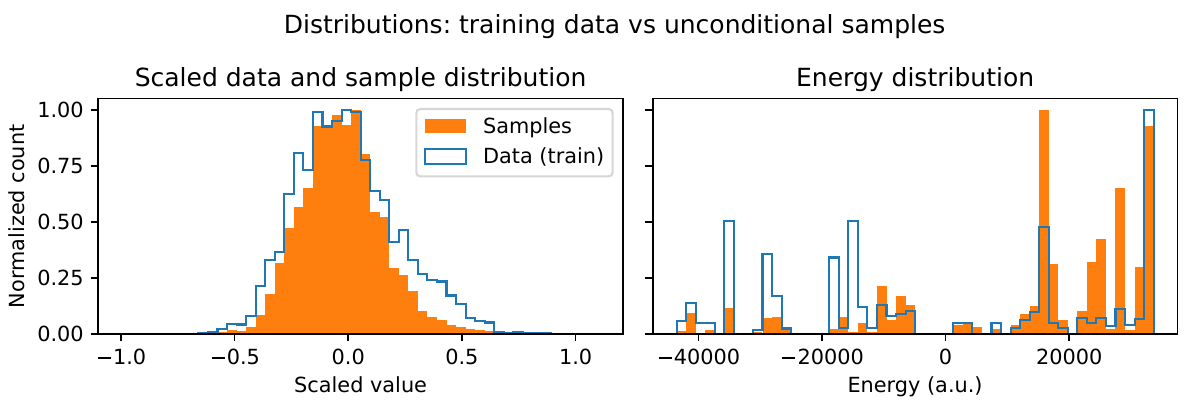}
	\caption[Histograms of scaled values and energy for unconditional samples]{\label{fig:uncond_energy}Left: all scaled inputs from the training set vs samples inputs. The distributions are similar, but the EBM does not appear to learn more extreme values. Right: corresponding energy distribution. The EBM learns all the modes, but the probability mass is not properly distributed.}
\end{figure}

The full unconditional distribution for each input (or at 16 ms for time-series data) can be seen in fig. \ref{fig:uncond_dist_full}. Notably, although most -- if not all -- modes of the distribution are covered, the mass associated with each mode may not agree. This behavior is evident in the aggregate energy distribution seen in fig. \ref{fig:uncond_energy}. On average, the unconditional samples have higher energy than the data. In terms of the scaled values of all of the inputs, the model appears to struggle with extreme values. This could hint towards the need for data augmentation so that chains can properly mix, or a need for a different random initialization before commencing Langevin dynamics.

\section{The replay buffer and configuring the sampler}

\begin{figure}
	\centering
	\includegraphics[width=300pt]{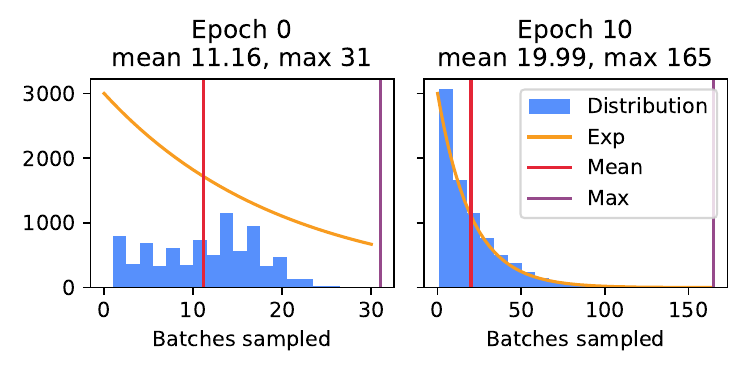}
	\caption[Distribution of batches in replay buffer samples]{\label{fig:buffer_distribution}Distribution of batches in replay buffer. When training is starting (epoch 0, left), the number of batches each sample experiences is low and somewhat uniform. After 10 training epochs (right), the number of batches experienced by a sample converged to an exponential distribution.}
\end{figure}

This replay fraction leads to a mean of 1/0.05=20 batch iterations for each chain, with half the chains experiencing $\ln(2)/0.05 \approx 14$ batches. The replay buffer requires 4 to 5 epochs to converge to an exponential distribution in the number of steps experienced by each chain in the buffer. This diversity of chain lengths likely encourages quick convergence of the chain but good long-term samples (on average, each chain experience 600 MCMC updates/steps). The distribution of the batches among the replay buffer can be seen in fig. \ref{fig:buffer_distribution}. One downside of this long buffer is that if a sample ends up in a region far outside the domain ($-1$ to 1) it can lead to collapse of the sampler and exploding gradients; samples have a $\approx$ 50\% chance of lasting 194 batches if one goes awry. Implementing a reject step for these prodigal samples may improve training stability. 

In general, if the model is not learning, increase the number of sample steps, decreasing the step size, and decreasing the learning rate are beneficial steps to take. A lower learning rate decreases the change per batch in the energy surface, so stale sample chains from the replay buffer may find themselves in more familiar territory than if the energy function is rapidly changing. More sample steps relaxes the requirement for the samples to rapidly find realistic (lower energy) locations, perhaps leading to shallower gradients and greater stability. Much work needs to be done to really understand the dynamics of the sampler in the training process. The samples from the replay buffer sampling may also benefit from a non-memoryless distribution so that the chains are guaranteed a certain number of sample steps over their lifetime, with a hard limit so that chains do not persist for too long. 

\begin{figure}
	\centering
	\includegraphics[width=\linewidth]{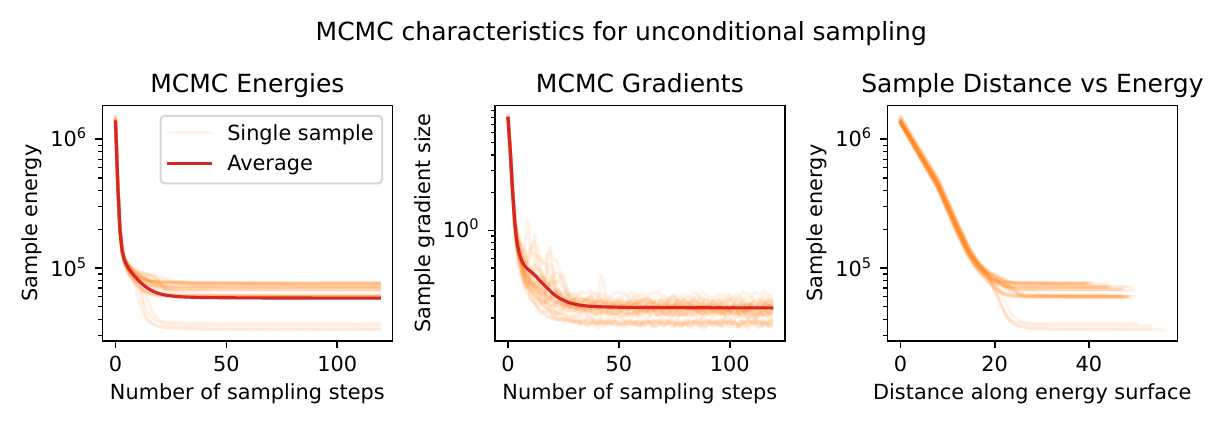}
	\caption[MCMC energies, gradients, and integrated trajectory length for unconditional samples]{\label{fig:uncond_mcmc}MCMC energies, gradients, and integrated trajectory length for unconditional samples. Left: the model converges after approximately 50 sampling steps. Middle: the gradients approach an asymptote; long-term samples are realistic. Right: integrated trajectory length show that individual MCMC trajectories vary in total distance traveled along the energy surface.}
\end{figure}

The MCMC trajectories of unconditional samples can be seen in fig. \ref{fig:uncond_mcmc}. Notably after a small number of sampling steps -- around 30 -- the energy surface gradients appear to flatten out and thus the energies of the samples level off. This leveling point may be determined by the number of samples steps used while training the model: each randomly-initialized sample runs for 30 steps before being added to the replay buffer. Despite converging relatively quickly, long-run MCMC chains look just as realistic as shorter-duration chains and do not exhibit the burn-in or high-saturation that has been observed in other work \cite{du_implicit_2020}. 

\end{document}